\documentclass[12pt]{article}

\usepackage[totalheight = 23cm, totalwidth = 17cm]{geometry}
\usepackage{amssymb,amsmath,amsfonts,amsbsy,graphicx,bm}

\usepackage{color,cancel,ulem}

\newcommand{\dis}[1]{\begin{equation}\begin{split}#1\end{split}\end{equation}}

\begin{document}

\begin{titlepage}

\begin{center}

{\LARGE \bf 
Brane/flux annihilation in double-throat system
}

\vskip 1.0cm

{\large
Min-Seok Seo$^{a}$ 
}

\vskip 0.5cm

{\it
$^{a}$Department of Physics Education, Korea National University of Education,
\\ 
Cheongju 28173, Republic of Korea
}

\vskip 1.2cm

\end{center}

\begin{abstract}

 We study the brane/flux annihilation in the double-throat system in which the corresponding cycles of two throats are    homologically equivalent.
 When we put $\overline{\rm D3}$-branes at only one of throat tips, two throats are no longer identical.
 Then the  brane/flux annihilation can be  interpreted as the process for reducing the difference between   two throat geometries through the transition to the supersymmetric vacuum.
 To see this, we describe the changes in the amounts of the NSNS 3-form flux supporting the B-cycle contained in two throat regions during the brane/flux annihilation.
 We also compare our analysis with the recently proposed thraxion scenario, which also describes the inequivalence of two throats caused by the different NSNS 3-form flux distributions in two throat regions.

\end{abstract}

\end{titlepage}

\newpage

\section{Introduction}

 Supersymmetry (SUSY) breaking in the fluxed background has drawn considerable interest in string model building as it is an essential ingredient for realizing  $\Lambda$CDM and   inflation models which well describe the current and  early universe, respectively.
 In Type IIB Calabi-Yau orientifold compactifications with fluxes,  SUSY is broken by  anti-D3 ($\overline{\rm D3}$)-branes at the tip of the Klebanov-Strassler (KS) throat. 
The backreaction of 3-form fluxes supporting the deformed conifold geometry generates the warp factor \cite{Klebanov:2000hb}, which redshifts the  SUSY breaking scale down to the low scale.
 This may  explain  the electroweak scale of $100$GeV if we allow some amount of fine tuning.

 As the SUSY breaking $\overline{\rm D3}$-brane state is not absolutely stable, it  eventually evolves into the supersymmetric vacuum through the brane/flux annihilation.
 More concretely, a number $p$ of   $\overline{\rm D3}$-branes polarize into an NS5-brane wrapping a contractible cycle along $S^3$ of the deformed conifold.
 This extracts the unit NSNS 3-form flux quantum and produces D3-branes, the number of which coincides with the RR 3-form flux quantum $M$ such that we are left with the D3-brane number $M-p$.
 When $p <M $, the transition into the supersymmetric vacuum is completed after a single step of the brane/flux annihilation.
 The transition rate  is extremely suppressed for $ p/M \lesssim 0.08$, in which case the SUSY breaking state persists for a sufficiently long enough time \cite{Kachru:2002gs}.
 Then the  anti-de Sitter (AdS) minimum can be uplifted to the  de Sitter (dS) metastable vacuum without spoiling the K\"ahler moduli stabilization induced by the quantum effects \cite{Kachru:2003aw, Balasubramanian:2005zx}, as well as the complex structure moduli and  axio-dilaton stabilization through the fluxes  \cite{Dasgupta:1999ss, Giddings:2001yu}.
  \footnote{Since the Gukov-Vafa-Witter (GVW) superpotential generated by fluxes linearly depends on the axio-dilaton \cite{Gukov:1999ya}, the axio-dilaton can easily be light and even become unstable through the mixing with the K\"ahler moduli \cite{Choi:2004sx} (see also \cite{Seo:2021kyi} for a recent discussion). } 
  While the SUSY breaking state is no longer metastable for $p/M > 0.08$, the transition to the supersymmetric vacuum in this case permits inflation \cite{DeWolfe:2004qx}.
  
  For $p >M $, the transition to the supersymmetric vacuum requires about $p/M$ steps of the brane/flux annihilation.
  During the transition, the vacuum energy density gradually  decreases as the NS5-brane oscillates between the north and south pole along $S^3$ \cite{Gautason:2016cyp, DiazDorronsoro:2017qre}.
 This indeed is one way to realize the  axion monodromy,  the unfolding of the  periodic axion field space induced by the coupling of an axion to fluxes \cite{Silverstein:2008sg, McAllister:2008hb} (for various  models, see, e.g., \cite{Kaloper:2008fb, Marchesano:2014mla, Blumenhagen:2014gta, Hebecker:2014eua, Ibanez:2014swa, Escobar:2015ckf}).
  The axion monodromy has been known as the natural model of the trans-Planckian inflation since the potential is protected against the quantum gravity corrections by the `hidden' axion periodicity : the discrete shift of the axion field can be compensated by the change in the flux quantum.
  However, the recent swampland conjectures (for reviews, see, e.g., \cite{Brennan:2017rbf, Palti:2019pca, vanBeest:2021lhn}) suggest that   the trans-Planckian inflation is in conflict with quantum gravity, which motivated intense scrutiny on the validity of the trans-Planckian axion monodromy from various directions, e.g., \cite{Ibanez:2015fcv, Hebecker:2015zss, Baume:2016psm, Brown:2016nqt, Klaewer:2016kiy, Valenzuela:2016yny, Blumenhagen:2017cxt, Buratti:2018xjt, Scalisi:2018eaz, Scalisi:2020jal}.
 Regarding the brane/flux annihilation, it turns out that  the  number of initial $\overline{\rm D3}$-branes cannot be arbitrarily large   to maintain the deformed conifold geometry against the backreaction \cite{Bena:2018fqc}, which restricts the geodesic distance traversed by   the NS5-brane position modulus to be sub-Planckian \cite{Scalisi:2020jal}.
  
  The geometry describing the brane/flux annihilation deviates from  Calabi-Yau 3-fold (CY$_3$) by the backreaction of $\overline{\rm D3}$-branes as well as fluxes.
  In particular, even though a number of throats are related in homology, that is, the corresponding cycles of throats belong to the same homology class, putting $\overline{\rm D3}$-branes at only one of  throat tips makes the geometry of the throat different from others, violating the CY$_3$ condition. 
 The purpose of this article is to investigate  the change in the geometry of such multi-throat system during the brane/flux annihilation  by taking the double-throat system as an example. 
 Intriguingly, the multi-throat system  has been considered to realize the axion monodromy \cite{Retolaza:2015sta, Hebecker:2015tzo, Hebecker:2018yxs, Carta:2021uwv}.
 A remarkable recent suggestion is to promote the 3-form fluxes accumulated at $S^2$s of the throat tips    to a complexified axion field, termed the thraxion \cite{Hebecker:2018yxs, Carta:2021uwv}.
For a nonzero  thraxion value, the amounts of 3-form fluxes contained in the respective throats become different.
Then throats related in homology are no longer identical such that the geometry deviates from CY$_3$.
 This indeed is  similar  to what happens during the brane/flux annihilation as putting $\overline{\rm D3}$-branes at one of throat tips also makes the amounts of flux contained in two throats different.

 This article is organized as follows. 
 In Sec. \ref{sec:doublethroat}, the geometry of the double-throat system  is described in detail.
 In Sec. \ref{sec:B/Fann}, we present the different distributions of the NSNS 3-form flux contained in two throats when $\overline{\rm D3}$-branes are located at only one of throat tips, which make the geometries of two throats different.
 We also study  the change in the geometry during the brane/flux annihilation to show that as the process proceeds two throats become less different.
 After comparing our results with the thraxion scenario in Sec. \ref{sec:discussion}, we conclude.
 We note that Sec. \ref{sec:RevThroat} and Sec. \ref{sec:RevAnn} are devoted to the reviews on the throat geometry and the brane/flux annihilation to summarize essential features needed for our discussion and settle the notation. 
 Readers who are familiar with these topics may skip them.

\section{Geometry of double-throat system}
\label{sec:doublethroat}

\subsection{Review on Klebanov-Strassler throat geometry}
\label{sec:RevThroat}

We begin with brief review on the geometry of the KS throat,  the warped deformed conifold \cite{Klebanov:2000hb} (for more details, see, e.g., \cite{Herzog:2001xk, Strassler:2005qs, Gwyn:2007qf}).
 The conifold  is the geometry around a conical singularity with a $T^{1,1}\sim S^3 \times S^2$ base.
 This can be `deformed' by smoothing out the $S^3$ singularity, which is described in terms of complex coordinates $w_i$ ($i=1,\cdots, 4$) as
 \dis{w_1^2 + w_2^2 + w_3^2 + w_4^2 =\epsilon^2,}
 where    $\epsilon$ is a parameter for the nonzero radius of $S^3$.
 When $S^2$ instead of $S^3$ singularity is smoothed out, we say the conifold is `resolved'.

 The background geometry of CY$_3$ $X$ containing the deformed conifold is supported by 3-form fluxes.
 To see this explicitly, consider a basis of the de Rham cohomology group $H^3 (X, \mathbb{Z})$ with generators $\alpha_I$ and $\beta^I$ ($I=0, \cdots h^{2,1}(X)$), the Poincar\'e dual to the homology basis $(A^I, B_I)$ of $H_3(X, \mathbb{Z})$ satisfying
 \dis{& A^I \cdot A^J=B_I \cdot B_J=0,\quad A^I \cdot B_J=\delta^I_J,
 \\
 &\int_{A^J}\alpha_I=-\int_{B_I}\beta^J=\int_X \alpha_I \wedge \beta^J=\delta_I^J.}
 Then  fluxes $F_3$ and $H_3$ are quantized as
 \footnote{Quantities we consider are written in the string unit by setting $\alpha'=\ell_s^2/(4\pi^2)=1$.}
\dis{F_3=4\pi^2 (M^I \alpha_I+M_I \beta^I)\quad\quad 
H_3=4\pi^2 (K^I \alpha_I+K_I \beta^I),}
which are constrained by the tadpole cancellation condition,
\dis{\frac{1}{(4\pi^2)^2}\int_X F_3 \wedge H_3=M^IK_I-M_I K^I =-N_{D_3}+\frac12 N_{O_3}+\frac{\chi(X)}{24},\label{eq:tadpole}}
where $\chi(X)$ is the Euler characteristic of CY$_4$ in the F-theory compactifications \cite{Sethi:1996es}.
For the conifold geometry, we take the flux quanta along the A-cycle ($S^3$) and the B-cycle ($S^2 \times \mathbb{R}$) to be $(M^z, M_z)=(M, 0)$ and $(K^z, K_z)=(0, K)$, respectively, i.e.,
\dis{\int_A F_3 = 4\pi^2 M,\quad\quad \int_B H_3 =-4\pi^2 K,\quad\quad \int_B F_3 = \int_A H_3 =0.
\label{eq:oflux}}
The fluxes backreact on the geometry to generate the warp factor $h(y)$ which depends only on $y$, the coordinate for $\mathbb{R}$ on the B-cycle,  such that the metric is given by
\footnote{The warped metric is often written as $ds^2=e^{2A}e^{-6u}g_{\mu\nu}dx^\mu dx^\nu+e^{-2A}e^{2 u}ds_6^2$ by defining $e^A=h^{-1/4}$ and $e^{4u}={\rm Im}\rho$.
Here $\rho$ is the K\"ahler modulus determining the overall internal volume.
}
 \dis{ds^2=h^{-1/2}(y) ({\rm Im}\rho)^{-3/2} g_{\mu\nu}dx^\mu dx^\nu + h^{1/2}(y)({\rm Im}\rho)^{1/2} ds_6^2.\label{eq:throatmetric}}
Near the tip of the throat, the metric of the deformed conifold is written with respect to the basis of 1-forms $\{g_i\}$ on $T^{1,1}$ as
 \dis{&ds_6^2=\frac12 \epsilon^{4/3}K(y)\Big[\frac{1}{3 K(y)^3}\big(dy^2+ g_5^2\big)+\cosh^2\Big(\frac{y}{2}\Big)\big(g_3^2+g_4^2\big)+\sinh^2\Big(\frac{y}{2}\Big)\big(g_1^2+g_2^2\big)\Big],
 \\
 &K(y)=\frac{\big(\sinh (2y)-2y\big)^{1/3}}{2^{1/3}\sinh y},\label{eq:tipgeometry}}
 and the warp factor is given by
 \dis{&h(y)=(g_s M)^2\frac{2^{2/3}}{\epsilon^{8/3}}I(y),
 \\
 &I(y)=\int^\infty_y dx \frac{x \coth x-1}{\sinh^2 x}\big(\sinh (2x)-2x\big)^{1/3}.\label{eq:warpf}}
 We note that the $\epsilon$-dependence in $ds_6^2$ is cancelled by that in $h^{1/2}$ such that the physical size of $S^3$ at the tip is independent of $\epsilon$.
 Using $K(0)=(2/3)^{1/3}$ and defining $b_0^2=(4/3)^{1/3}I(0)^{1/2}$ ($I(0)\simeq 0.71805$), the metric is approximated as 
 \dis{ds^2=&\Big(\frac{2^{1/3}}{3^{1/3} b_0^2}\Big)\frac{\epsilon^{4/3}}{g_sM}({\rm Im}\rho)^{-3/2} g_{\mu\nu}dx^\mu dx^\nu 
 \\
 &+ b_0^2(g_sM)({\rm Im}\rho)^{1/2}\Big[\frac14 dy^2+d\psi^2+\sin^2\psi(d\omega^2+\sin^2\omega d\varphi^2)+\frac{y^2}{4}(d\tilde{\omega}^2+\sin^2\tilde{\omega}d\tilde{\varphi}^2)\Big].\label{eq:tipmetric}}
 Here, $g_i$ are parametrized with respect to $S^3$ coordinates $(\psi, \omega, \varphi)$ and  $S^2$ coordinates $(\tilde{\omega}, \tilde{\varphi})$, which are appropriate to describe the brane/flux annihilation  (for derivation, see, e.g., \cite{Nguyen:2019syc}).

 Meanwhile, the throat geometry far from the tip   is close to AdS$_5 \times$T$^{1,1}$, the metric of which is written as 
\dis{ds^2 =\frac{r^2}{R(r)^2}({\rm Im}\rho)^{-3/2}g_{\mu\nu}dx^\mu dx^\nu+\frac{R(r)^2}{r^2}({\rm Im}\rho)^{1/2}\big(dr^2+r^2 ds_{{T}^{1,1}}^2\big), \label{eq:farmetric}}
where $dr^2+r^2 ds_{{T}^{1,1}}^2$ is the $y > 1$ limit of $ds_6^2$ in \eqref{eq:tipgeometry}, with $r$ defined as
\dis{r^2=\frac{3}{2^{5/3}} \epsilon^{4/3}e^{2y/3}.\label{eq:r/tau}}
We note that the tip is located at $r_{\rm IR}=r(y=0)=(2^{-5/3}\times 3)^{1/2}\epsilon^{2/3}$ which is nonzero for the deformed conifold ($\epsilon \ne 0$).
In the warp factor $h(r)\simeq R(r)^4/r^4$, $R(r)$ is interpreted as the AdS$_5$ size.
More explicitly \cite{Klebanov:2000nc}, 
\dis{&R(r)^4=\frac{27\pi }{4}g_sM\Big(k(r)+\frac{3 b_0^4 }{4\pi}(g_s M)\Big),\label{eq:AdSsize}}
where $k(r)$ is  the NSNS 3-form flux contained in the region $[r_{\rm IR}, r]$ :
 \dis{k(r)=-\frac{1}{4\pi^2}\int_{S^2(r)\times [r_{\rm IR}, r]} H_3=-\frac{1}{4\pi^2}\int_{S^2(r)} B_2= \frac{3 g_s M}{2\pi}\log\Big(\frac{r}{r_{\rm IR}}\Big).\label{eq:k-r}}
 Whereas the B-cycle in general extends beyond the throat region, when we focus on the physics in the throat, it is typical to assume that all the  flux is contained in the throat region only.
 That is, given $r_{\rm UV}$,  the $r$ coordinate of the UV end of the throat (hence the throat region extends over $[r_{\rm IR}, r_{\rm UV}]$),  $k(r_{\rm UV})$ is nothing more than $K$.
 On the contrary, in this article, we will consider the double-throat system, in which the flux  is distributed over two throats as well as the bulk region belonging to the B-cycle.

In Type IIB Calabi-Yau orientifold compactifications, the complex structure moduli  including one that determines the size of $\epsilon$  are stabilized by turning on 3-form fluxes, as explicitly shown in \cite{Giddings:2001yu}. 
 In the effective supergravity description,  the flux-induced superpotential called the Gukov-Vafa-Witten (GVW) superpotential is given by \cite{Gukov:1999ya}
  \dis{W_{\rm GVW}=  a \int_X \Omega \wedge G_3,\label{eq:GVW}}
where $\Omega$ is the holomorphic 3-form, $G_3= F_3-\tau H_3$ with $\tau=C_0+i e^{-\phi}$ being an axio-dilaton, and $a$ is an ${\cal O}(1)$ constant in the string unit.
Here the holomorphic 3-form $\Omega$ is written with respect to $A$- and $B$-periods of $\Omega$,
 \dis{Z^I=\int_{A^I}\Omega,\quad\quad {\cal F}_I=\int_{B_I}\Omega}
 as
\dis{\Omega=Z^I \alpha_I-{\cal F}_I \beta^I.\label{eq:Omega}}
In addition, we can define ${\cal F}(Z^I)$ satisfying ${\cal F}_I=\partial_I {\cal F}$ from which the prepotential $F(t^a)$ is given by ${\cal F}(Z^I)=(Z^0)^2F(t^a)$ with $t^a=Z^a/Z^0$ ($a=1, \cdots h^{2,1}(X)$).
Then the GVW superpotential is written as
\dis{W_{\rm GVW}=a\Big[(Z^I M_I+F_I M^I) -\tau (Z^I K_I + F_I K^I)\Big].\label{eq:GVWflux}}
Meanwhile, the K\"ahler potential for the complex structure moduli in CY$_3$ is written as
 \dis{K_{\rm cs}=-\log\Big(i \int_X \Omega\wedge \overline{\Omega}\Big)=-\log\Big(i|Z^0|^2\Big[2(F-\overline{F})-(t^a-\overline{t}^a)(F_a+\overline{F}_a)\Big]\Big),\label{eq:Kahler0}}
 whereas the warping modifies $\Omega\wedge \overline{\Omega}$  to $h \Omega\wedge \overline{\Omega}$ \cite{DeWolfe:2002nn}.

 The deformation parameter $\epsilon^2$ which controls the size of $S^3$ is proportional to the stabilized value of the complex structure modulus $z$ associated with the A- and  B-cycle of the throat.
 More precisely, the equation of motion
 \dis{\tilde{\nabla}^2 h^{-1}=h^{-1/2}\frac{G_{mnp}\overline{G}^{mnp}}{12{\rm Im}\tau}+\cdots,}
 where $\tilde{\nabla}^2$ being the Laplacian with respect to the unwarped CY$_3$ metric $ds_6^2$ indicates that  under $ds_6^2 \to \lambda ds_6^2$, the warp factor $h$ for the throat given by \eqref{eq:warpf} ($\propto \epsilon^{-8/3}$)  scales as $h \to  \lambda^{-2} h$ while the overall volume $V_X =({\rm Im}\rho)^{3/2}$ scales as $V_X \to \lambda^3V_X$.
% \footnote{The scaling behavior of $V_X$ comes from
%  \dis{V_X=\int d^6y \sqrt{\tilde{g}}e^{-4A_{\rm tot}}.}
% Here $e^{-4A_{\rm tot}}$ is  the warp factor for the {\it overall internal volume}, which should not be confused with the warp factor for the {\it throat} $e^{-4A_0}$.}
This implies the relation $\epsilon^{8/3}={\rm Im}\rho |z|^{4/3}$ where $\rho$ is the K\"ahler modulus determining the overall CY$_3$ volume \cite{Giddings:2005ff}.

 The A- and   B-period for the throat geometry are given by
\dis{\int_A \Omega =z \quad\quad \int_B \Omega ={\cal F}_z=\frac{z}{2\pi i}\log z +({\rm holomorphic}).\label{eq:zcycle}}
Taking 3-form flux quanta in \eqref{eq:oflux} into account, and setting $Z^0=1$, one finds that the leading terms of the K\"ahler potential and the GVW superpotential in the limit of $|z| \to 0$ are given by
\dis{& K_{\rm tot}=-\log\Big(\frac{|z|^2}{2\pi}\log|z|^2 +{\rm constant}+\cdots\Big)-\log[-i(\tau-\overline{\tau})]-3\log[-i(\rho-\overline{\rho})],
\\
&W_{\rm GVW}=a\Big[\frac{M}{2\pi i}z\log z -K \tau z +\cdots\Big],\label{eq:K/W}}
for small warping.
When the warping is strong, the modification of the K\"ahler potential gives \cite{Douglas:2007tu, Douglas:2008jx}
\dis{G_{z{\overline z}}=\Big(i\int_X h\Omega\wedge\overline{\Omega}\Big)^{-1}\Big[c_0+c_1\log z+c_2\frac{(g_sM)^2}{({\rm Im}\rho) |z|^{4/3}}\Big],}
for ${\cal O}(1)$ coefficients $c_0, c_1, c_2$.
Since it is dominated by the last term for $|z| \ll 1$,   the potential for $|z|$ is written as
\dis{V_{\rm KS}=\frac{g_s}{16 c_2 ({\rm Im}\rho)^2}\frac{|z|^{4/3}}{(g_s M)^2}\Big|\frac{M}{2\pi}\log|z|+\frac{K}{g_s}\Big|^2,\label{eq:KSpotential}}
 where the axio-dilaton is stabilized at $\tau = i/g_s$.
Regardless of the warping, at the minimum of the potential,  $z$ is stabilized at an exponentially small value,
\dis{|z| \simeq |z_0| {\rm exp}\Big(-\frac{2\pi K}{g_s M }\Big),}
  satisfying $0=D_z W_{\rm GVW}\simeq \frac{M}{2\pi i}\log z-i\frac{K}{g_s}$ as well as $D_{\tau} W_{\rm GVW}=0$.
  Here an ${\cal O}(1)$ coefficient $z_0$ comprehensively parametrizes the effects from sub-dominant terms in $K_{\rm tot}$ and $W_{\rm GVW}$.

\subsection{Description for double-throat system}
\label{sec:double-throat-default}

%%%%%%%%%%%%%%%%%%%%%%%%%%%%%%%%%%%%%%%%%%%%%%%%%%%%%%%%%%%%%%%%%%%%%%%%%%%%%%%%%%%%%%%%
 \begin{figure}[!ht]
  \begin{center}
   \includegraphics[width=0.8\textwidth]{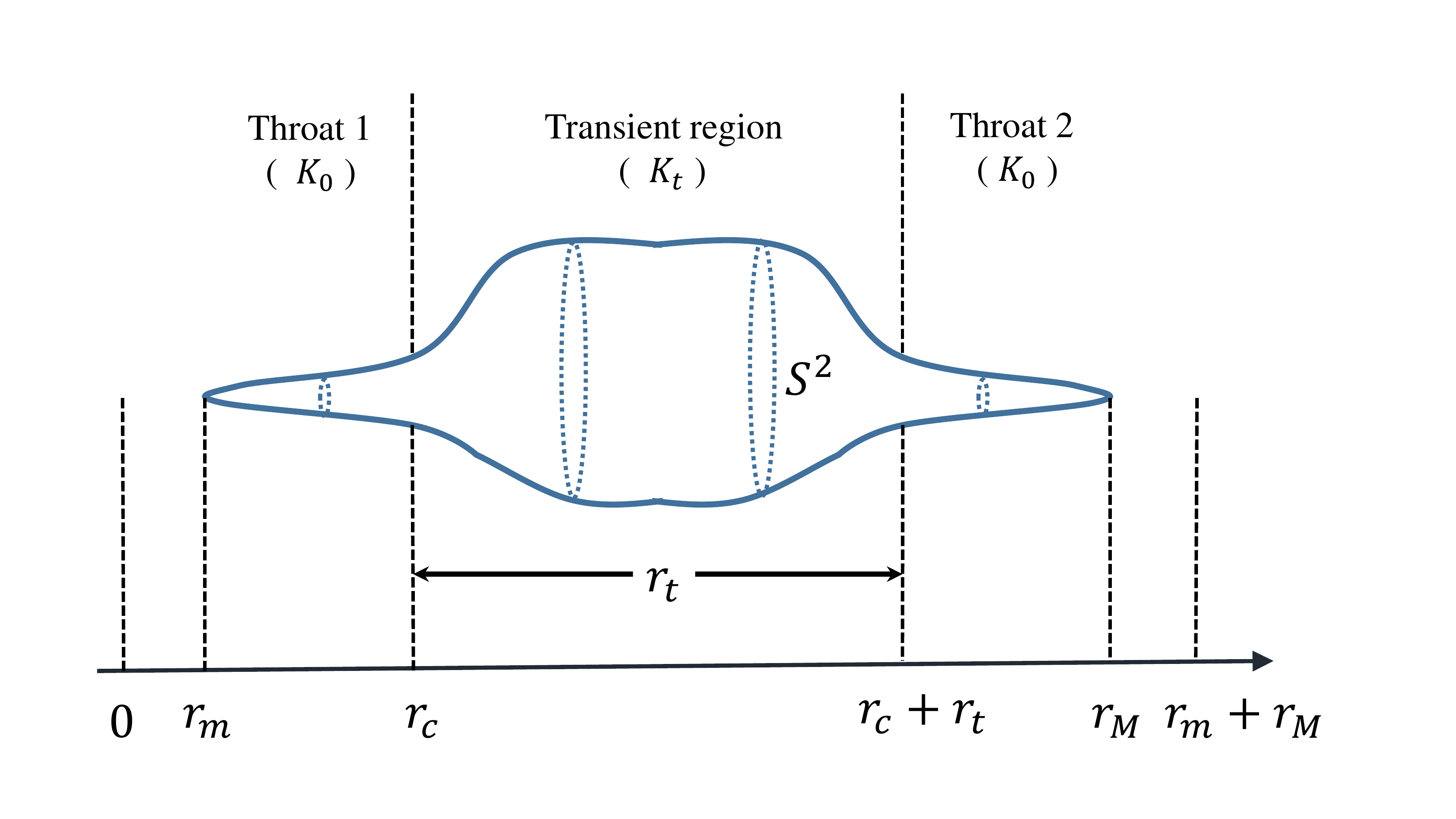}
  \end{center}
 \caption{Schematic shape of the B-cycle extending over throat 1, transient region, and throat 2 along the $r$ coordinate. 
 Contributions to the NSNS 3-form flux quantum from each region are given by $K_0$ (throat 1), $K_t$ (transient region), and $K_0$ (throat 2), respectively, satisfying $K=2 K_0+K_t$.
  }
\label{fig:throat1}
\end{figure}
%%%%%%%%%%%%%%%%%%%%%%%%%%%%%%%%%%%%%%%%%%%%%%%%%%%%%%%%%%%%%%%%%%%%%%

We now consider the double-throat system, the CY$_3$ containing  two throats related in homology.
In this case,  $A^1$ and $A^2$, the A-cycles associated with two throats,  are homologically equivalent $[A^1]=[A^2]$ in the sense of $A^1 = A^2 +\partial a_4$ for some 4-chain $a_4$, and  intersect with a single B-cycle satisfying $A^1 \cdot B=A^2 \cdot B=1$.
The B-cycle is topologically equivalent to $S^3$, or  a family of $S^2$s,  extending over two throats as well as a part of the bulk.
In the throat regions, $S^2$  is identified with $S^2$ of $T^{1,1}$, which collapses at the tip.
A part of the bulk belonging to the B-cycle which we will call  the `transient region'  connects two throats smoothly.

Since the flux quanta associated with $A^1$ and $A^2$ are the same, we set
\dis{&\int_{A^1} F_3 = \int_{A^2} F_3= 4\pi^2 M,\quad\quad \int_B H_3 =-4\pi^2 K,
\\
&\int_{A^1} H_3 = \int_{A^2} H_3=  0,\quad\quad \int_B F_3 =  0.\label{eq:fluxes}}
Here,  the NSNS 3-form flux spreads over the whole B-cycle to give  $K$.
In order to describe this more systematically, we introduce a `radial coordinate' $r$ perpendicular to $S^2$ on the B-cycle.
 It runs over the first throat (`throat 1') for $r_m \leq r < r_{c}$, over the transient region for $r_{c} \leq r < r_{c}+r_t$, and over the second throat (`throat 2') for $ r_{c}+r_t \leq r \leq r_M$, as depicted in Fig. \ref{fig:throat1}.
 For throat 1 (throat 2), the locations of the tip and the UV end are given by $r_m$ ($r_M$) and $r_c$ ($r_c+r_t$), respectively.
  Moreover, the $r$ coordinate  is taken   to coincide with   $r$ defined in \eqref{eq:r/tau} in the throat regions.
   If $\epsilon$, or equivalently, $z$ were zero, the tips of two throats would be located at $0$ and $r_m+r_M$, respectively.
   Since $\epsilon \ne 0$,   $r_m$, the position of the tip of throat 1, is identified with $r_{\rm IR}$, which deviates from zero  as can be read off from \eqref{eq:r/tau} : $r_m = (2^{-5/3}\times 3)^{1/2}\epsilon^{2/3}$.
 For the same reason, the tip of   throat 2 is located at $r_M$ instead of $r_m+r_M$.

 When throat 1 and throat 2 have the same distribution of branes and fluxes,  the geometries of two throats are identical.
 Since the  lengths of two throats in this case must coincide,  
 \dis{&r_c - r_m = r_M - (r_c+r_t) \label{eq:rel1}}
is satisfied.
Moreover, two throats contain the same amount of the NSNS 3-form flux,   denoted by $K_0$.
Given the contribution $K_t$ from the transient region, the total flux quantum is written as  $K= 2K_0+K_t$.
The distribution of the NSNS 3-form flux  in the throat regions can be written  as a function of $r$ using \eqref{eq:k-r}.
Whereas \eqref{eq:k-r} is an expression valid far from the tip ($y > 1$), it is more or less a good estimation since $S^2$  shrinks to the zero size at the tip such that the flux cannot be accumulated there.
We will return to  this issue in Sec. \ref{sec:discussion}, where the comparison with the resolved throat is discussed.  
  For   throat 1, the amount of   flux  accumulated from the tip to $r$  is simply given by \eqref{eq:k-r}, with $r_{\rm IR}$ replaced by $r_m$.
  Then the flux $K_0$ contained in throat 1 satisfies 
  \dis{K_0=\frac{3g_s M}{2\pi}\log\Big(\frac{r_c}{r_m}\Big).\label{eq:throat1flux}}
  For   throat 2, as opposed to  throat 1, $r$ increases as we approach the tip. 
  Noting that $r$ in \eqref{eq:k-r} is the distance from $r=0$, the position of the tip in the absence of deformation ($\epsilon=0$), we find   the corresponding distance for throat 2 to be $r_m+r_M-r$ since $r_m+r_M$ is the position of the tip of   throat 2 for $\epsilon=0$.   
  The distance from $r_m+r_M$ to the actual ($\epsilon \ne 0$) tip is $r_m+r_M - r_M =r_m$ as expected, and that to the UV end   is $r_m+r_M-(r_c+r_t)$, which is the same as $r_c$ as can be inferred from  \eqref{eq:rel1}.  
  Then the amount of  flux accumulated in the throat 2 region $[r_c+r_t, r]$  is given by
  \dis{k_{\rm throat 2}(r)&=K_0 - k(r_m+r_M-r)
  =K_0-\frac{3 g_s M}{2\pi}\log\Big(\frac{r_m+r_M-r}{r_m}\Big)
  \\
% &=K_0- \frac{3 g_s M}{2\pi}\Big[\log\Big(\frac{r_c}{r_m}\Big)+\log\Big(\frac{r_m+r_M-r}{r_c}\Big)\Big]
% \\
 &=\frac{3 g_s M}{2\pi}\log\Big(\frac{r_c}{r_m+r_M -r}\Big)=\frac{3 g_s M}{2\pi}\log\Big(\frac{r_c}{2r_c+r_t -r}\Big).}
Here $k(r_m+r_M-r)$ in the first line is the value of  \eqref{eq:k-r} with $r$ replaced by the distance  $r_m+r_M-r$.
In the last equality, the relation $r_m+r_M = 2r_c+r_t $   obtained from \eqref{eq:rel1}  is used.
It is straightforward to check that $k_{\rm throat 2}(r_c+r_t)=0$ and $k_{\rm throat 2}(r_M)=K_0$.
Meanwhile, the amount of flux accumulated from $r_c$ to $r$ in the transient region,   denoted by $k_t(r)$, depends on details of  the bulk geometry, with constraints $k_t(r_c)=0$ and $k_t(r_c+r_t)=K_t$.
In summary, the accumulation of the NSNS 3-form flux along $r$ in the B-cycle is given by
\begin{equation}
\label{eq:horwave}
k(r) =
\left\{
\begin{array}{ll}
\frac{3 g_s M}{2\pi}\log\big(\frac{r}{r_m}\big) & \text{for} \quad r_m \leq r <r_c
\vspace{0.5em}
\\
K_0 + k_t(r) & \text{for} \quad r_c \leq r <r_c+r_t
\vspace{0.5em}
\\
K_0+K_t+\frac{3 g_s M}{2\pi}\log\big(\frac{r_c}{2r_c+r_t -r}\big) & \text{for} \quad r_c+r_t \leq r \leq r_M
\end{array}
\right.
\, 
.
\end{equation}

%While $z$ parametrizes the size of deformed $S^3$ at the tip, the cancellation between $z$ dependence in $ds^6$ and the warp factor  the warp factor at the vicinity of tip given by \eqref{eq:warpf} is exponentially suppressed such that $h^{-1/4}(r_{\rm min})\propto z^{1/3}\sim {\rm exp}[-\frac{2\pi}{3g_s}\frac{K}{M}]$,
%it does not directly appear in the geometry since the $z$ dependence in the warp factor and $ds_6^2$ are cancelled with each other.
%The physically meaningful warp effect is the ratio of the warp factor at the tip to that at the bulk.
The warp factor is exponentially suppressed in the throat regions by the backreaction of fluxes supporting the deformed conifold.
Thus we expect $h\simeq 1$ in the bulk which includes the UV end  of the throat.
Then the relative warping of   throat 1 compared to the bulk  is given by
\dis{\frac{h^{-1/4}(r_m)}{h^{-1/4}(r_c)}\sim \frac{r_m}{r_c}=e^{-\frac{2\pi K_0}{3 g_s M}},}
as \eqref{eq:throat1flux} indicates.
Since throat 2 has the same geometry as   throat 1,   the relative warping of throat 2 is identical to that of throat 1.
We also note from \eqref{eq:AdSsize} that the AdS$_5$ size at the UV end of the throat is given by
\dis{R^4=\frac{27\pi }{4}g_s M\Big(K_0+\frac{3 b_0^4 }{4\pi}(g_s M)\Big),}
for both throat 1 and 2.
The AdS$_5\times$T$^{1,1}$ geometry around the UV end is smoothly connected to the geometry of the transient region at boundary, $r=r_c$ for   throat 1 and $r=r_c+r_t$ for  throat 2.

\section{Brane/flux annihilation in double-throat system}
\label{sec:B/Fann}

 \subsection{Review on brane/flux annihilation}
 \label{sec:RevAnn}
 
The SUSY breaking state is realized by putting  $\overline{\rm D3}$-branes at the tip of the throat.
It is unstable, so evolves toward the supersymmetric vacuum through the brane/flux annihilation (see \cite{Gautason:2015tla} for the discussion on  generic features).
More concretely, when a number $p$ of $\overline{\rm D3}$-branes are located at the north pole of the A-cycle  $S^3$, $\psi=0$ in terms of the coordinates in \eqref{eq:tipmetric}, they  polarize  into an NS5-brane, which  eventually produces $M$ D3-branes at the south pole, $\psi=\pi$, by reducing the unit NSNS 3-form flux quantum ($K \to K-1$).
Then the total D3-brane number at the tip changes from $-p$ to $-p+M$ such that the state becomes more supersymmetric hence more stable.
The transition preserves the tadpole cancellation condition \eqref{eq:tadpole} as
\dis{MK -p+\cdots = M(K-1)+ (-p+M)+\cdots,}
where $\cdots$ contains contributions from fluxes or  branes which are not relevant to the brane/flux annihilation so remains unchanged during the process.
The NS5-brane  fills the non-compact spacetime and wraps a trivial cycle along $\omega, \varphi$ directions of the A-cycle.
Then   the induced metric on the NS5-brane is given by
\dis{ds_{\rm NS5}^2=&\Big(\frac{2^{1/3}}{3^{1/3} b_0^2}\Big)\frac{\epsilon^{4/3}}{g_sM}({\rm Im}\rho)^{-3/2} g_{\mu\nu}dx^\mu dx^\nu + b_0^2(g_sM)({\rm Im}\rho)^{1/2}\sin^2\psi(d\omega^2+\sin^2\omega d\varphi^2).}
The potential describing the brane/flux annihilation is obtained  from the NS5-brane action
\dis{-S_{\rm NS5}=\frac{T_5}{g_s^2}\int d^6\xi \sqrt{-g}\sqrt{{\rm det}(g_{2-{\rm cycle}}+g_s 2\pi{\cal F}_2)}+ T_5\int B_6,}
where $2\pi{\cal F}_2=2\pi F_2+C_2$ and $T_5=1/(2\pi)^5=T_3/(2\pi)^2$.
Using the imaginary self-duality of $G_3$ ($\star_6G_3=i G_3$ or $\star_6 H_3=-g_s F_3$ which will be directly applied to the second Chern-Simons term) at the minimum, as well as
\dis{&C_2=M\Big(\psi-\frac12\sin 2\psi\Big)\sin\omega d\omega\wedge d\varphi+\cdots,
\\
&2\pi\int_{S^2}F_2=-4\pi^2 p,}
 we obtain
 \footnote{As noted in \cite{Kachru:2002gs}, the term proportional to $\pi(p/M)$ is added to make $V_{\rm NS5}=0$ at the supersymmetric minimum.}
\dis{V_{\rm NS5}(\psi)=&\frac{T_3}{g_s}\frac{M}{\pi}\Big[\frac{2^{2/3}}{3^{2/3}b_0^4}\frac{\epsilon^{8/3}}{(g_sM)^2}\frac{1}{({\rm Im}\rho)^3}\Big]
\\
&\times\Big\{\sqrt{b_0^4({\rm Im}\rho) \sin^4\psi+\Big(\frac{\pi p}{M}-\big(\psi-\frac12\sin 2\psi\big)\Big)^2}+\Big(\frac{\pi p}{M}-\big(\psi-\frac12\sin 2\psi\big)\Big)\Big\}.}
We note that the term in the square brackets in the first line is nothing more than  $h^{-1}$ coming from $\sqrt{-g}$, and redshifts the SUSY breaking scale.
At $\psi=0$, the potential is reduced to that generated by $p$ $\overline{\rm D3}$-branes,
\dis{V_{\overline{\rm D3}}=&\frac{T_3}{g_s}\Big[\frac{2^{2/3}}{3^{2/3}b_0^4}\frac{\epsilon^{8/3}}{(g_sM)^2}\frac{1}{({\rm Im}\rho)^3}\Big]\times(2p) .\label{eq:aD3pot}}
Meanwhile, at $\psi=\pi$, the potential becomes
\dis{V_{\rm NS5}(\pi)=&\frac{T_3}{g_s} \Big[\frac{2^{2/3}}{3^{2/3}b_0^4}\frac{\epsilon^{8/3}}{(g_sM)^2}\frac{1}{({\rm Im}\rho)^3}\Big]\times\Big\{ \big| p-M\big|+ (p-M)\Big\}.}
When $p < M$, $V_{\rm NS5}(\pi)=0$ thus the transition to the supersymmetric vacuum is completed after a single step of the brane/flux annihilation.
In contrast, for $p  >M$,  the number of D3-branes produced in the single step of the brane/flux annihilation is not sufficiently large enough to eliminate all $\overline{\rm D3}$-branes, and $V_{\rm NS5}(\pi)$ is given by \eqref{eq:aD3pot} with $p$ replaced by $p-M$.
Since the SUSY breaking $\overline{\rm D3}$-branes remain at the tip, the state is still unstable.
Then the brane/flux annihilation takes place once more : a number $p-M$ of $\overline{\rm D3}$-branes polarize into an NS5 brane to produce another $M$ D3-branes at $\psi=2\pi$, which in fact is the same position as $\psi=0$, the north pole.
As a result of this additional process, the NSNS 3-form flux quantum is  reduced to $K-2$ and the D3-brane number becomes $2M-p$.
In this way, $\lceil p/M \rceil$ steps   of  the brane/flux annihilation occur  until the state reaches the supersymmetric minimum.
\footnote{Here $\lceil x \rceil$ is the ceiling function, given by the least integer larger than or equal to $x$.
Meanwhile, $\lfloor x \rfloor$ is the floor function, given by the largest integer less than or equal to $x$.
Thus, for a non-integer $x$,  $\lceil x \rceil = \lfloor x \rfloor + 1$.} 
During the whole process, the potential decreases to zero as $\psi$ oscillates between the north pole ($0, 2\pi, 4\pi, \cdots$) and  south pole ($\pi, 3\pi, \cdots$) as $0 \to \pi \to 2\pi \to 3\pi \to \cdots$.
Such a non-periodic behavior of the potential for the `axion' $\psi$ is a typical feature of the axion monodromy.
 While the final state has the NSNS 3-form flux quantum $K-\lceil p/M \rceil$  and the D3-brane number $-p+\lceil p/M \rceil M$ (especially, $0$ for an integer $p/M$), the tadpole cancellation condition \eqref{eq:tadpole} remains unchanged.

It is remarkable that the $\overline{\rm D3}$-brane potential \eqref{eq:aD3pot} depends on the complex structure modulus $z$ through $\epsilon^{8/3}={\rm Im}\rho |z|^{4/3}$.
The total potential $V_{\rm KS}+V_{\overline{\rm D3}}$ where $V_{\rm KS}$ is given by \eqref{eq:KSpotential} is minimized at
\dis{|z|=|z_0|{\rm exp}\Big[-\frac{2\pi K}{g_s M }-\frac34+\sqrt{\frac{9}{16}-\frac{2^{14/3}c_2}{\pi 3^{2/3}b_0^4 a^2}\frac{p}{(g_sM)^2}}\Big].\label{eq:zmodified}}
As pointed out in \cite{Bena:2018fqc}, this shows that $|z|$ cannot be stabilized at the nonzero value if $p \gg (g_s M)^2$ such that the term  in the square root becomes negative.
This, however, does not  exclude the case of $p > M$ completely as $g_s M\gg 1$ is required by the condition of the considerable size of the tip in the string unit for the validity of the supergravity description (see the prefactor of $ds^6$ in \eqref{eq:tipmetric}).
During the brane/flux annihilation, the $\overline{\rm D3}$-brane number,  the minus of the D3-brane number may be replaced by \cite{Scalisi:2020jal}
\dis{v(\psi)=\frac{M}{2\pi}\Big\{\sqrt{b_0^4({\rm Im}\rho) \sin^4\psi+\Big(\frac{\pi p}{M}-\big(\psi-\frac12\sin 2\psi\big)\Big)^2}+\Big(\frac{\pi p}{M}-\big(\psi-\frac12\sin 2\psi\big)\Big)\Big\},\label{eq:v(psi)}}
provided $v(\psi)$ is positive.
If the $\overline{\rm D3}$-brane number becomes negative, $v\simeq 0$ through the cancellation between, roughly speaking, the negative $\pi (p/M)-\psi$ and its absolute value. 
Moreover, $\lceil (p-v(\psi))/M \rceil$, roughly $\psi/\pi$ is interpreted as the number of oscillations promoted to the continuous real number, which runs over $[0, \lceil p/M \rceil]$.
From the positivity of the term in the square root in \eqref{eq:zmodified}, the excursion range of $\psi$, or the number of oscillations is constrained as \cite{Scalisi:2020jal} 
\dis{\Delta \psi=\pi\Big\lceil \frac{p}{M} \Big\rceil  \lesssim \frac{9\pi}{16}\Big(\frac{\pi 3^{2/3}b_0^4 a^2}{2^{14/3}c_2}\Big)(g_s^2M).}

%%%%%%%%%%%%%%%%%%%%%%%%%%%%%%%%%%%%%%%%%%%%%%%%%%%%%%%%%%%%%%%%%%%%%%%%%%%%%%%%%%%%%%%%
 \begin{figure}[!t]
  \begin{center}
   \includegraphics[width=0.8\textwidth]{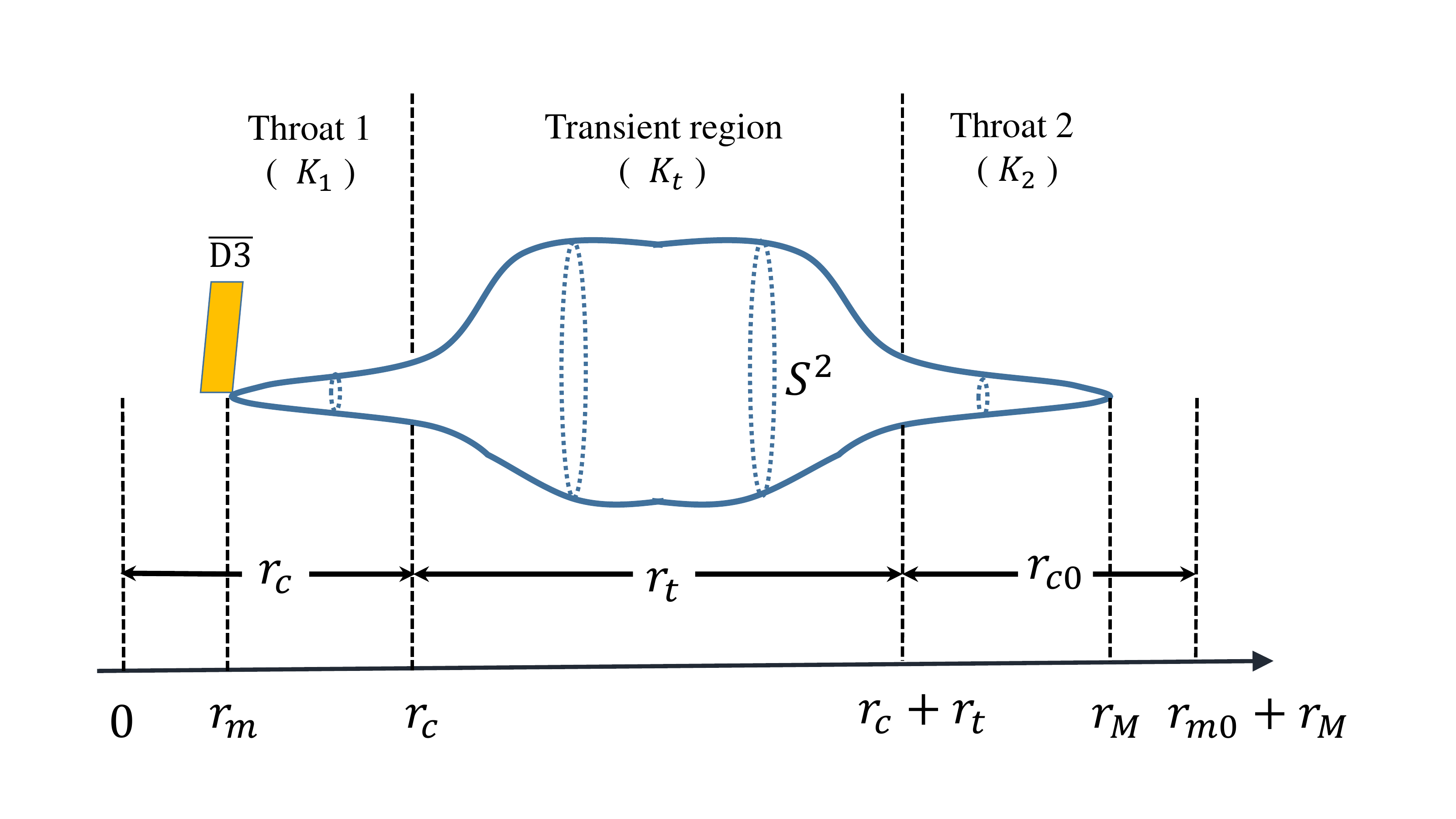}
  \end{center}
 \caption{Schematic shape of the B-cycle along the $r$ coordinate in the presence of $\overline{\rm D3}$-branes at the tip of throat 1.
 The backreaction of $\overline{\rm D3}$-branes and the different  flux distributions in two throats make two throat geometries different.
 Contributions to the NSNS 3-form flux quantum from each region are given by $K_1$ (throat 1), $K_t$ (transient region), and $K_2$ (throat 2), respectively, satisfying $K=2 K_0+K_t$ and $K_1=K_2+p/M$.
  }
\label{fig:throat2}
\end{figure}
%%%%%%%%%%%%%%%%%%%%%%%%%%%%%%%%%%%%%%%%%%%%%%%%%%%%%%%%%%%%%%%%%%%%%%

\subsection{Geometry of double-throat system with $\overline{\rm D3}$-branes}
\label{Sec:geo-double-throat}

 We now move onto the double-throat system.
 When we put $p$ $\overline{\rm D3}$-branes at the tip of  throat 1 only, the whole geometry deviates from the warped CY$_3$, as  throat 1 is no longer identical to throat 2. 
Moreover, the flux contained in   throat 1,  $K_1$, is different from that in  throat 2,   $K_2$.
The tadpole cancellation condition \eqref{eq:tadpole} in this case is written as
\dis{MK=M(K_1+K_t+K_2)=p+\cdots,}
where $\cdots$ contains contributions from  fluxes and branes other than $p$ $\overline{\rm D3}$-branes, and $K_t$ is the flux contained in the transient region as  discussed in Sec. \ref{sec:doublethroat}.
Then we claim that $K_1$ and $K_2$ are related as $K_1=K_2 +(p/M)$.
It is consistent with the fact that the dynamics  associated with $\overline{\rm D3}$-branes like the brane/flux annihilation  is local to the tip of   throat 1, as will be clear in the following discussion.
Indeed, if $p/M$ is an integer, the brane/flux annihilation reduces the $\overline{\rm D3}$-branes completely such that the CY$_3$ geometry satisfying $K_1=K_2$ is recovered.

Just as in Sec. \ref{sec:doublethroat}, we consider the radial coordinate $r$ which runs over $r_m \leq r < r_c$ for   throat 1, $r_c \leq r < r_c+r_t$ for the transient region, and $r_c+r_t \leq r \leq r_M$ for   throat 2, as can be found in Fig. \ref{fig:throat2}.
Even though the notations are reused, their explicit values are different.
First, the $\overline{\rm D3}$-brane potential \eqref{eq:aD3pot} is added to the potential for $z_{\rm throat1}$ only.
Then  $z_{\rm throat1}$ is modified to \eqref{eq:zmodified}, which changes $r_m$, the position of the  tip measured from $r=0$ (the position of the tip for $\epsilon=0$, i.e., in the absence of deformation), through \eqref{eq:r/tau}.
In contrast, since $\overline{\rm D3}$-branes do not backreact on the tip of    throat 2, $z_{\rm throat2}$ is not modified by the $\overline{\rm D3}$-brane potential.
Thus the position of the tip of  throat 2 for $\epsilon=0$  can be written as $r_M+r_{m0}$, in which $r_{m0}$ satisfies
\dis{\frac{r_{m0}}{r_m}=\Big|\frac{z_{\rm throat2}}{z_{\rm throat1}}\Big|^{1/3}={\rm exp}\Big[\frac14-\frac13\sqrt{\frac{9}{16}-\frac{2^{14/3}c_2}{\pi 3^{2/3}b_0^4 a^2}\frac{p}{(g_sM)^2}}\Big].\label{eq:zratio}}
This ratio is practically one for $p/M\ll 1$, for instance, $p/M\lesssim 0.08$ as considered in \cite{Kachru:2002gs} which realizes the metastable SUSY breaking state.
Since we are interested in the relative size of two throats, we consider  $r_{c0}\equiv r_{m0}+r_M-(r_c+r_t)$, the distance between $r_c+r_t$, the position of the UV end of throat 2, and $r_M+r_{m0}$, the position of the tip of throat 2 for $\epsilon=0$.
If $\overline{\rm D3}$-branes were not present, $r_{c0}$ would be the same as $r_c$.
From $\frac{3g_sM}{2\pi}\log(r_c/r_m)=K_1$ and $\frac{3g_sM}{2\pi}\log(r_{c0}/r_{m0})=K_2$, the condition $K_1=K_2+p/M$ reads
\dis{\frac{3 g_sM}{2\pi}\log\Big(\frac{r_c}{r_m}\Big)=\frac{3 g_sM}{2\pi}\log\Big(\frac{r_{c0}}{r_{m0}}\Big)+\frac{p}{M},\label{eq:ridenold}}
or equivalently,
\dis{r_c%&=\Big(\frac{r_m}{r_{m0}}\Big)e^{\frac{2\pi}{3g_sM}\frac{p}{M}}( r_{m0}+r_M-(r_c+r_t))
= \Big(\frac{r_m}{r_{m0}}\Big)e^{\frac{2\pi}{3g_sM}\frac{p}{M}}r_{c0}. \label{eq:ridentity}}
That is, $r_c$ is enhanced by the factor ${\rm exp}[\frac{2\pi}{3g_sM}\frac{p}{M}]$ compared to $r_{c0}$ as throat 1 contains $p/M$ more NSNS 3-form flux than throat 2.
The additional factor $r_m/r_{m0}$ given by \eqref{eq:zratio} reflects the backreaction of $\overline{\rm D3}$-branes on the tip of throat 1 only.
Therefore, the amount of flux accumulated in the throat 2 region $[r_c+r_t, r]$  becomes
\dis{k_{\rm throat2}(r)&=K_2-\frac{3g_sM}{2\pi}\log\Big(\frac{r_{m0}+r_M-r}{r_{m0}}\Big)=\frac{3g_sM}{2\pi}\log\Big(\frac{r_{c0}}{r_{m0}+r_M-r}\Big)
\\
&=\frac{3g_sM}{2\pi}\log\Big(\frac{r_{c0}}{r_{c0}+r_c+r_t-r}\Big),\label{eq:kthroat2}}
which satisfies $k_{\rm throat2}(r_c+r_t)=0$ and $k_{\rm throat2}(r_M)=K_2$.
 From our discussions so far, the accumulation of the NSNS 3-form flux  in the B-cycle along $r$ is summarized as
\begin{equation}
\label{eq:fluxsumm}
k(r) =
\left\{
\begin{array}{ll}
\frac{3 g_s M}{2\pi}\log\big(\frac{r}{r_m}\big) & \text{for} \quad r_m \leq r <r_c
\vspace{0.5em}
\\
K_1 + k_t(r) & \text{for} \quad r_c \leq r <r_c+r_t
\vspace{0.5em}
\\
K_1+K_t+\frac{3g_sM}{2\pi}\log\Big(\frac{r_{c0}}{r_{c0}+r_c+r_t-r}\Big) & \text{for} \quad r_c+r_t \leq r \leq r_M
\end{array}
\right.
\, 
.
\end{equation}

\subsection{Dynamics of double-throat system in brane/flux annihilation}
\label{sec:dthroatb/f}

After a single step of the brane/flux annihilation, the NSNS 3-form flux quantum changes from $K$ to $K-1$, while the D3-brane number changes from $-p$ to $-p+M$.
Since the polarization of $\overline{\rm D3}$-branes inducing the brane/flux annihilation is the local process at the tip of   throat 1, we expect that the reduction of the NSNS 3-form flux quantum by one unit comes from the change in the flux contained in throat 1, $K_1 \to K_1-1$ while $K_t$ and $K_2$ remain intact.
This indeed is consistent with the relation $K_1=K_2+(p/M)$ : $K_1 \to K_1-1$ under $p\to p-M$ and $K_2 \to K_2$.
 At the same time, $|z_{\rm throat1}|$ is   changed from \eqref{eq:zmodified} to
 \dis{|z^{(1)}_{\rm throat1}|=|z_0|{\rm exp}\Big[-\frac{2\pi (K-1)}{g_s M }-\frac34+\sqrt{\frac{9}{16}-\frac{2^{14/3}c_2}{\pi 3^{2/3}b_0^4 a^2}\frac{(p-M)}{(g_sM)^2}}\Big],}
 whereas $|z_{\rm throat2}|={\rm exp}[-\frac{2\pi}{g_sM}K]$ is changed to $|z^{(1)}_{\rm throat2}|={\rm exp}[-\frac{2\pi}{g_sM}(K-1)]$.

In throat 1,   the change $ z_{\rm throat1} \to z^{(1)}_{\rm throat1}$ leads to the scaling of $r$ in  throat 1 to $r^{(1)}$ through \eqref{eq:r/tau} as
 \dis{&r^{(1)}=e^{\frac{2\pi}{3g_sM}+\frac13\gamma(p, p-M)}r,
 \\
 &\gamma(p, p-M)=\sqrt{\frac{9}{16}-\frac{2^{14/3}c_2}{3^{2/3}b_0^4 a^2}\frac{(p-M)}{(g_sM)^2}}-\sqrt{\frac{9}{16}-\frac{2^{14/3}c_2}{\pi 3^{2/3}b_0^4 a^2}\frac{p}{(g_sM)^2}},}
 from which the position of the tip of throat 1  is changed into $r^{(1)}_m$ in the same way, 
 \dis{r^{(1)}_m=e^{\frac{2\pi}{3g_sM}+\frac13\gamma(p, p-M)}r_m.}
Meanwhile, the position of the UV end of throat 1 after a single step of the brane/flux annihilation is determined by  the reduction $K_1 \to K_1-1$ of the flux contained in throat 1,  
\dis{r^{(1)}_c=e^{\frac{2\pi}{3g_sM}(K_1-1)}r^{(1)}_m = e^{\frac13\gamma(p, p-M)}r_c.\label{eq:rcstep1}}
After $\lceil p/M \rceil$  steps of the brane/flux annihilation, the transition to the supersymmetric vacuum is completed, and the positions of the tip and the UV end of throat 1 are given by
\dis{&r^{f}_{m} = e^{\frac{2\pi}{3g_sM}\big\lceil \frac{p}{M} \big\rceil+\frac13\gamma\big(p, p-\big\lceil \frac{p}{M} \big\rceil M\big)} r_m,
\\
&r^{f}_{c} =  e^{\frac13\gamma\big(p, p-\big\lceil \frac{p}{M} \big\rceil M\big)} r_c.}
%We note that $r^{(1)}$ in   throat 1 contains a factor ${\rm exp}[(1/3)\gamma (p, p-M)]$ which is larger than one and originated from  the change in $z_{\rm throat1}$ by the $\overline{\rm D3}$-brane potential, \eqref{eq:aD3pot}.
%Then we may rescale $r^{(1)}$ such that $\overline{r}^{(1)}=e^{-\frac13\gamma(p, p-M)}r^{(1)}$, in terms of which $\overline{r}^{(1)}_m\simeq {\rm exp}[(2\pi)/(3g_s M)] r_m$ and $\overline{r}^{(1)}_c \simeq r_c$.
%This shows the shrink of the first throat as a result of the brane/flux annihilation more explicitly.

On the other hand, in throat 2, $ z_{\rm throat2} \to z^{(1)}_{\rm throat2}$ after a single step of the brane/flux annihilation results in
\dis{r^{(1)}=e^{\frac{2\pi}{3g_sM}}r,}
according to which $r_{m0}$ and $r_M$ are changed into
\dis{r^{(1)}_{m0}=e^{\frac{2\pi}{3g_sM}}r_{m0},\quad
r^{(1)}_M=e^{\frac{2\pi}{3g_sM}}r_M,\label{eq:rmMstep1}}
respectively.
As $r$, $r_{m0}$, and $r_M$ are  scaled by the same factor, $k_{\rm throat2}(r^{(1)})=k_{\rm throat2}(r)$ is satisfied, as can be found in the first equality in \eqref{eq:kthroat2}.
 One direct consequence is that $K_2$ remains  unchanged under the brane/flux annihilation, which requires
\dis{r^{(1)}_{c0}=e^{\frac{2\pi}{3g_sM}}r_{c0}.\label{eq:rc0step1}}
 Then from \eqref{eq:rcstep1}, \eqref{eq:rmMstep1}, and \eqref{eq:rc0step1}, we find that % together with the changes in the position of the tip and the UV end of  throat 1 and $p\to p-M$, \eqref{eq:ridentity} indicates that 
  the length of the transient region becomes
\dis{r^{(1)}_t&=r^{(1)}_{m0}+r^{(1)}_M-r^{(1)}_c-r^{(1)}_{c0}
\\
%&=e^{\frac{2\pi}{3g_sM}}r_{m0} + e^{\frac{2\pi}{3g_sM}}r_M-\Big[e^{\frac13\gamma(p, p-M)}+e^{\frac{2\pi}{3g_sM}}e^{-\frac{2\pi}{3g_sM}\frac{p}{M}}\Big(\frac{r_{m0}}{r_m}\Big)\Big]r_c
%\\
&=e^{\frac{2\pi}{3g_sM}} r_t+\Big[e^{\frac{2\pi}{3g_sM}}-e^{\frac13\gamma(p, p-M)}\Big]r_c.}
 After the transition to the supersymmetric vacuum is completed, we obtain
\dis{&r^{f}_{m0} = e^{\frac{2\pi}{3g_sM}\big\lceil \frac{p}{M} \big\rceil} r_{m0},
\quad r^{f}_M = e^{\frac{2\pi}{3g_sM}\big\lceil \frac{p}{M} \big\rceil} r_M,
\quad r^{f}_{c0} = e^{\frac{2\pi}{3g_sM}\big\lceil \frac{p}{M} \big\rceil} r_{c0},
\\
&r^{f}_t=e^{\frac{2\pi}{3g_sM}\big\lceil \frac{p}{M} \big\rceil}  r_t +
\Big[e^{\frac{2\pi}{3g_sM}\big\lceil \frac{p}{M} \big\rceil}-e^{\frac13\gamma\big(p, p-\big\lceil \frac{p}{M} \big\rceil M\big)}\Big] r_c.}
 This clearly shows that the length of throat 2, $r_{c0}^f-r_{m0}^f$, or $r^f_{M}-r^f_{c}-r^f_{t}$ is scaled by the factor ${\rm exp}[\frac{2\pi}{3g_sM}\big\lceil \frac{p}{M} \big\rceil]$ compared to the original length $r_{c0}-r_{m0}$.

As discussed around \eqref{eq:v(psi)}, during the brane/flux annihilation, the $\overline{\rm D3}$-brane number can be promoted to $v(\psi)$ provided it is positive while   the number of oscillations of $\psi$  can be promoted to $\lceil (p-v(\psi))/M \rceil$.
From this,  the stabilized values of the $z$ modulus at two throat tips are written as
\dis{&|z^\psi_{\rm throat1}|={\rm exp}\Big[-\frac{2\pi}{g_sM}\big(K-\Big\lceil \frac{1}{M}(p-v(\psi))\Big\rceil\big)-\gamma(v(\psi), 0)\Big],
\\
&|z^\psi_{\rm throat2}|={\rm exp}\Big[-\frac{2\pi}{g_sM}\big(K-\Big\lceil \frac{1}{M}(p-v(\psi))\Big\rceil \big)\Big],\label{eq:z1and2}}
whereas various quantities relevant to the throat geometry are given by  
\dis{&r^{\psi}_{m} = e^{\frac{2\pi}{3g_sM}\big\lceil \frac{1}{M}(p-v(\psi))\big\rceil+\frac13\gamma (p, v(\psi) )} r_m,
\quad
r^{\psi}_{c} =  e^{\frac13\gamma(p, v(\psi) )} r_c \label{eq:throat1coor}}
for   throat 1, and
 
\dis{&r^{\psi}_{m0} = e^{\frac{2\pi}{3g_sM}\big\lceil \frac{1}{M}(p-v(\psi))\big\rceil} r_{m0},
\quad r^{\psi}_{M} = e^{\frac{2\pi}{3g_sM}\big\lceil \frac{1}{M}(p-v(\psi))\big\rceil} r_{M},
\quad r^{\psi}_{c0} = e^{\frac{2\pi}{3g_sM}\big\lceil \frac{1}{M}(p-v(\psi))\big\rceil} r_{c0},
\\
&r^{\psi}_t=e^{\frac{2\pi}{3g_sM}\big\lceil \frac{1}{M}(p-v(\psi))\big\rceil}  r_t +
\Big[e^{\frac{2\pi}{3g_sM}\big\lceil \frac{1}{M}(p-v(\psi))\big\rceil}-e^{\frac13\gamma(p, v(\psi))}\Big] r_c \label{eq:throat2coor}}
for   throat 2.
 We remind the reader that $r_m$ and $r_{m0}$ are related by \eqref{eq:zratio}, or $r_{m0}={\rm exp}[\frac13\gamma(p,0)]r_m$.
In addition, $r_c$ and $r_{c0}$ are related by \eqref{eq:ridentity}.

 Our analysis so far shows how the  change in $z$ (or equivalently, $\epsilon$) affects  the throat geometry through the change in the  $r$ coordinate, even though the $z$ dependence in $ds_6^2$ is cancelled by that in the warp factor. 
Especially, the relative size of two throats changes during the brane/flux annihilation.
From the length of throat 1,
\dis{r_c^\psi-r_m^\psi = \Big(e^{\frac{2\pi}{3g_sM}\big(K_1-\big\lceil \frac{1}{M}(p-v(\psi))\big\rceil\big)} -1\Big) e^{\frac{2\pi}{3g_sM}\big\lceil \frac{1}{M}(p-v(\psi))\big\rceil +\frac13\gamma (p, v(\psi) )} r_m ,}
and that of throat 2,
\dis{r_{c0}^\psi-r_{m0}^\psi &=  \Big(e^{\frac{2\pi}{3g_sM}K_2} -1\Big) e^{\frac{2\pi}{3g_sM}\big\lceil \frac{1}{M}(p-v(\psi))\big\rceil} r_{m0} 
\\
&=\Big(e^{\frac{2\pi}{3g_sM}K_2} -1\Big) e^{\frac{2\pi}{3g_sM}\big\lceil \frac{1}{M}(p-v(\psi))\big\rceil+\frac13\gamma (p, 0 )} r_m ,}
 the ratio between two throat lengths is given by
\dis{\frac{r_{c0}^\psi-r_{m0}^\psi}{r_c^\psi-r_m^\psi}=\Big[\frac{e^{\frac{2\pi}{3g_sM}K_2} -1}{e^{\frac{2\pi}{3g_sM}\big(K_1-\big\lceil \frac{1}{M}(p-v(\psi))\big\rceil\big)} -1}\Big]e^{\frac13\gamma (v(\psi), 0 )}.}
%Since $K_1=K_2+p/M$, the term in the square bracket in RHS is smaller than one until the last step.
%In the supersymmetric vacuum, this term becomes one for an integer $p/M$ as $\psi/\pi=p/M$, and larger than but close to one for a non-integer $p/M$ as $\psi/\pi=[p/M]+1$.
Since $v(\psi) \leq p$, the term in the square brackets is smaller than one until the final stage, while the second factor, ${\rm exp}[\frac13 \gamma(v(\psi), 0)]$ is larger than one but may be close to one so long as $p \ll (g_s M)^2$.
In any case, if $p/M$ is an integer, the ratio between two throat lengths at the final stage of the brane/flux annihilation becomes one : $v(\psi)=0$ at $\psi=p/M$ which gives $K_1-\lceil (p-v(\psi))/M\rceil=K_1-(p/M)=K_2$ whereas $\gamma(0,0)=0$.
In other words, through the brane/flux annihilation, the NSNS 3-form flux contained in throat 1, initially given by $K_1$, gets smaller until it coincides with $K_2$ and all the $\overline{\rm D3}$-branes at the tip of throat 1 disappear.
Then we can interpret the brane/flux annihilation as the process for restoring the equivalence of two throats as required by the CY$_3$ condition.
For a non-integer $p/M$, the state reaches the final supersymmetric vacuum at $\psi=\lfloor p/M \rfloor+1$, i.e., $v(\psi=\lfloor p/M\rfloor+1)=0$.
The D3-brane number  at the tip of throat 1 at the final stage is $-p+(\lfloor p/M \rfloor+1)M>0$, which means that there still remain D3-branes at the supersymmetric minimum.
Then the amounts of  NSNS 3-form flux contained in two throats are still different. 
Even though the geometries of two throats become less different through the brane/flux annihilation, since no more process reducing the D3-brane takes place in throat 1 at the final stage,   the lengths of two throats are still different hence the final geometry  still deviates from the warped CY$_3$ : 
\dis{\frac{r_{c0}^f-r_{m0}^f}{r_c^f-r_m^f}=\Big[\frac{e^{\frac{2\pi}{3g_sM}K_2} -1}{e^{\frac{2\pi}{3g_sM}\big(K_2+\frac{p}{M}-\big\lfloor\frac{p}{M}\big\rfloor-1\big)} -1}\Big]e^{\frac13\gamma(p-(\lfloor \frac{p}{M}\rfloor+1)M,0)}.}
We note that for the negative $\overline{\rm D3}$-brane number, it cannot be identified with $v(\psi)$ any longer as $v(\psi)$ becomes zero.
This is reflected in the last factor, which is smaller than one. 
Meanwhile, the term in the square brackets is larger than one.
It is dominant when $\gamma(p-(\lfloor \frac{p}{M}\rfloor+1)M,0)$ is sufficiently close to zero, in which case the length of  throat 1 is shorter than that of   throat 2.

 The AdS radii of two throats during the brane/flux annihilation are, according to \eqref{eq:AdSsize}, 
  \dis{&R_{\rm throat1}(r^\psi)^4=\frac{27\pi g_sM}{4}\Big(k(r^\psi)+\frac{3 b_0^4 }{4\pi}(g_s M)\Big),
 \\
 &R_{\rm throat2}(r^\psi)^4=\frac{27\pi g_sM}{4}\Big(K-k(r^\psi)+\frac{3 b_0^4 }{4\pi}(g_s M)\Big),}
  where $k(r)$ is given by \eqref{eq:fluxsumm}. 
 At the UV ends of two throats, the AdS radii are given by
 \dis{&R_{\rm throat1}^4=\frac{27\pi g_sM}{4}\Big(K_1-\Big\lceil \frac{1}{M}(p-v(\psi))\Big\rceil+\frac{3 b_0^4 }{4\pi}(g_s M)\Big),
 \\
 &R_{\rm throat2}^4=\frac{27\pi g_sM}{4}\Big(K_2+\frac{3 b_0^4 }{4\pi}(g_s M)\Big).}
 As expected, $R_{\rm throat2}$ remains unchanged whereas $R_{\rm throat1}$ decreases during the process, satisfying $R_{\rm throat1}  > R_{\rm throat2} $  until the final stage. 
 At the final supersymmetric vacuum,  two throat sizes coincide for an integer $p/M$, whereas the size of   throat 1 becomes  smaller than that of  throat 2 for a non-integer $p/M$ as $K_1-\lceil (p-v(\psi))/M\rceil=K_2+(p/M)-(\lfloor p/M\rfloor +1)<K_2$.
 Since the AdS$_5\times$T$^{1,1}$ geometries of two throats are smoothly connected to two boundaries of the transient region, the geometry of the transient region is deformed as well.

\section{Comparison with thraxion scenario}
\label{sec:discussion}

 We first summarize the qualitative features of the brane/flux annihilation in the double-throat system we have discussed.
 Since the corresponding cycles of two throats in the double-throat system are homologically equivalent, two throats are identical in terms of the (warped) CY$_3$ geometry.
By putting additional branes at one of the throat tips, say, the tip of   throat 1, two throats contain the different amounts of the NSNS 3-form flux and  the geometry   deviates from the warped CY$_3$.
% If we put   D3-branes, such deviation is just another vacuum solution at which energy is minimized and supersymmetric.
 In particular,  $\overline{\rm D3}$-branes at the tip make the state unstable as they break SUSY,  thus the transition into more stable supersymmetric vacuum through the brane/flux annihilation takes place.
 If $p/M$ is an integer, neither $\overline{\rm D3}$- nor D3-branes remain at the tip of   throat 1 in the final supersymmetric vacuum.
 Then the warped CY$_3$ is recovered, with the NSNS 3-from flux quantum  associated with the B-cycle being decreased by $ p/M $. 
 For a non-integer $p/M$, D3-branes remain at the tip of   throat 1, which means that the geometry of the supersymmetric vacuum  still deviates from the warped CY$_3$.
 In any case, the relative size of two throats changes during the brane/flux annihilation. 
 As the brane/flux annihilation is characterized by the position of the  NS5-brane along $S^3$ of the tip of throat 1, all dynamic quantities during the brane/flux annihilation like the D3-brane number as well as  the   NSNS 3-from flux quantum are expressed as  functions of $\psi$, the NS5-brane position modulus.
 The decrease in energy along the direction of $\psi$ reveals the axion monodromy property of $\psi$.

 The similar geometry can be found in the thraxion scenario \cite{Hebecker:2018yxs}.
  To see this, we first revisit the double-throat system without any additional $\overline{\rm D3}$-branes at the throat tips, as discussed in Sec. \ref{sec:double-throat-default}.
 In this case, the CY$_3$ structure imposes that two throats are identical, hence $z_1=z_2$ is satisfied.
  We now consider $B(r)$, the subregion of the B-cycle extending from the tip of   throat 1 to some value of $r$.
 Then $S^2(r)$, the section of the B-cycle at $r$,  becomes the boundary of $B(r)$ and the NSNS 3-form flux accumulated from the tip of throat 1 to $r$ can be written as
 \dis{k(r)=-\frac{1}{4\pi^2}\int_{B(r)} H_3 = -\frac{1}{4\pi^2}\int_{S^2(r)} B_2, }
 such that $k(r_M)=K$. 
 In the same way, we can consider the RR 3-form flux accumulated from the tip of throat 1 to $r$,
 \dis{m'(r)=-\frac{1}{4\pi^2}\int_{B(r)} F_3 = -\frac{1}{4\pi^2}\int_{S^2(r)} C_2.}
 Here we take $C_0=0$ such that the axio-dilaton is stabilized at $\tau=i/g_s$ and $F_3=dC_2-C_0 dB_2$ is just given by $dC_2$.
 Whereas the RR 3-form flux is distributed over the whole region of the B-cycle, including two throats and the transient region,   the total flux  vanishes in our flux assignment, i.e., $-4\pi^2 m'(r_M) = \int_B F_3=0$.
 Since $H_3$ and $F_3$ appear in the Type IIB effective supergravity action through the combination $G_3 = F_3 -(i/g_s)H_3$, we expect that the dynamics depends on  $k(r)$ and $m'(r)$ through the combined form, $g(r)=m'(r)-(i/g_s) k(r)$.
 In the warped deformed conifold, $S^2$ shrinks to the zero size at the throat tip, so the flux is not accumulated there, i.e., $g(r_m)=(-i/g_s)K-g(r_M-\delta)=0$ and $k(r)$ is just given by \eqref{eq:horwave}.
 \footnote{Here, $\delta$ is an infinitesimal number. 
 Since $g(r_M-\delta)$ represents the fluxes accumulated {\it from the tip of throat 1 to $r_M-\delta$} and the total flux quanta of the NSNS and RR 3-form are given by $k(r_M)=K$ and $m'(r_M)=0$, respectively, $(-i/g_s)K-g(r_M-\delta)$ represents the fluxes accumulated {\it at} $S^2$ of the tip of throat 2. }

 Meanwhile, \cite{Hebecker:2018yxs} suggested to  promote $g(r)$ to a dynamical field (in fact, the K\"ahler modulus), allowing $g(r_m)$ and $(-i/g_s)K-g(r_M-\delta)$ to have nonzero values. 
Then $S^2$s at two tips  no longer shrink but have finite sizes  and indeed, $g(r_m)=-[(-i/g_s)K-g(r_M-\delta)]$ provided $z_{\rm throat1}$ and $z_{\rm throat2}$ are nonzero.
This can be seen by rewriting the GVW superpotential as \cite{Hebecker:2018yxs}
\dis{W_{\rm GVW}=a\Big[\sum_{i=1}^2 \Big(\frac{M}{2\pi i}z_i \log z_i -\frac{i}{g_s}Kz_i\Big)+g(r_m)(z_{\rm throat1}-z_{\rm throat1})\Big], }
such that the stabilized values of  $z_{\rm throat1,2}$ are given by
\dis{&z_{\rm throat1}=z_0{\rm exp}\Big[-\frac{2\pi}{g_s M }(K+i g_s g(r_m))\Big]=z_0{\rm exp}\Big[-\frac{2\pi}{g_s M }(K+k(r_m))-\frac{2\pi i}{M}m'(r_m)\Big],
\\
&z_{\rm throat2}=z_0{\rm exp}\Big[-\frac{2\pi}{g_s M }(K- i g_s g(r_m))\Big]=z_0{\rm exp}\Big[-\frac{2\pi}{g_s M }(K-k(r_m))+\frac{2\pi i}{M}m'(r_m)\Big].\label{eq:thraxzs}}
If $g(r_m)$ is non-dynamical, it is just a Lagrange multiplier associated with the constraint $z_{\rm throat1}=z_{\rm throat2}\ne 0$ for the deformed conifold, which in turn imposes $ g(r_m)=0 $.
On the other hand, when we promote $g(r_m)$ to a dynamical field, the K\"ahler potential and the additional superpotential for $g(r_m)$ need to be taken into account.
Then \eqref{eq:thraxzs} indicates that the equations of motion are no longer simply solved by $z_{\rm throat1}=z_{\rm throat2}\ne 0$ as $g(r_m)$ may have the nonzero value.
Then the terms $K+k(r_m)$ and $K-k(r_m)$ in \eqref{eq:thraxzs} are interpreted as  a reflection of the following flux distribution : the NSNS 3-form flux is given by $k(r_m)$ at the tip of   throat 1 and the flux $K$ is additionally accumulated in the   B-cycle.
But the total flux quantum is $K$, rather than $K+k(r_m)$ since the flux $-k(r_m)$ is assigned at the tip of   throat 2.

The equivalence of two throats imposed by the  CY$_3$ condition is satisfied for either $g(r_m)=0$ or $g(r_m)\ne 0$ but $z_{\rm throat1}=z_{\rm throat2}=0$. 
The former case, $g(r_m)=0$ corresponds to the deformed throat we have discussed, in which the $S^3$ singularities are smoothed out by nonzero $z_{\rm throat1,2}$ whereas  $S^2$s   shrink to the zero size at  throat tips.
Here the non-dynamical $g(r_m)$ is regarded as a heavy field which is integrated out so does not appear in the low energy spectrum, leaving zero vacuum expectation value (vev).
 \footnote{However, its mass vanishes in the limit of $z_{\rm throat1,2} \to 0$.
 This property in fact is reflected in the K\"ahler potential in \eqref{eq:K/W} which is logarithmically dependent on $z_{\rm throat1,2}$ \cite{Strominger:1995cz}. }
Meanwhile, $z_{\rm throat1,2}$ in the low energy spectrum is stabilized  by the potential generated by  fluxes, satisfying $z_{\rm throat1}=z_{\rm throat2}\ne 0$.
In contrast, in the latter case, $z_{\rm throat1}=z_{\rm throat2}=0$ indicates that $S^3$s at throat tips shrink to the zero size.
Instead, as implied by $g(r_m)\ne 0$, the nonzero flux is accumulated at two throat tips, hence $S^2$s there have the finite size.
Then the throat geometry is given by the warped resolved conifold.
We note that unlike the case in which both $g(r_m)$ and $z_0$ are nonzero as considered in \eqref{eq:thraxzs}, the equivalence of two throats  in this case imposes $g(r_m)=+[(-i/g_s)K-g(r_M-\delta)]$, %i.e., the same amount of flux is accumulated at two tips, 
rather than `the same magnitude but the opposite sign'.
This is achieved when $z_{\rm throat1,2}$ are integrated out  leaving zero vevs, while  $g(r_m)$ in the low energy spectrum is stabilized at nonzero value.
In this way, the roles of $z$ (complex structure modulus) and $g(r_m)$ (K\"ahler modulus) are interchanged in the deformed  and   resolved conifold. 
Comparing these two geometries, the low energy spectrum of the resolved throat has one less complex structure modulus ($z$) but one more K\"ahler modulus ($g(r_m)$) than the deformed throat, and they are connected through the conifold transition \cite{Greene:1995hu}.

%%%%%%%%%%%%%%%%%%%%%%%%%%%%%%%%%%%%%%%%%%%%%%%%%%%%%%%%%%%%%%%%%%%%%%%%%%%%%%%%%%%%%%%%
 \begin{figure}[!t]
  \begin{center}
   \includegraphics[width=0.8\textwidth]{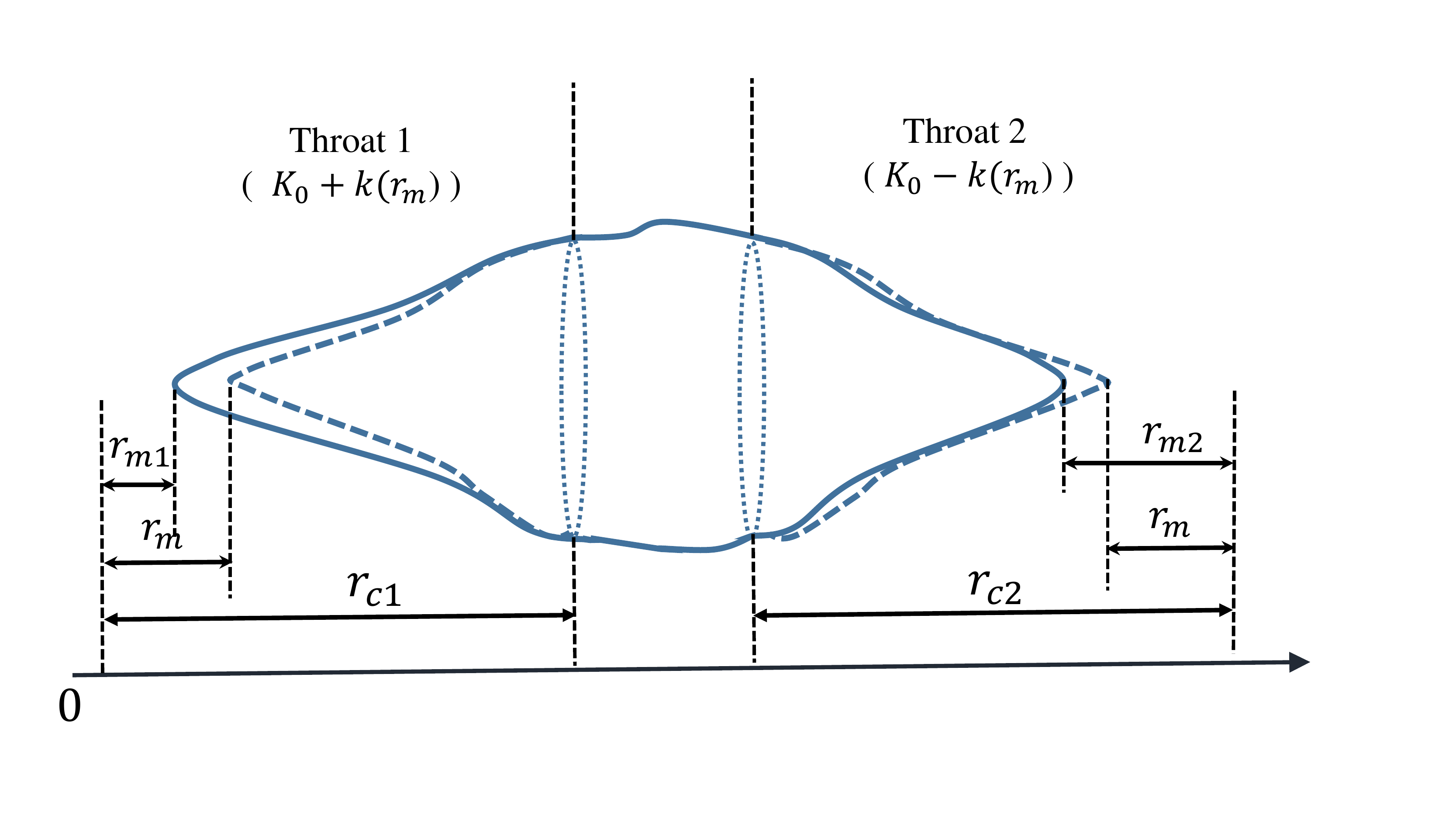}
  \end{center}
 \caption{Change in the throat geometry for nonzero $k(r_m)$ in the thraxion scenario. 
 The dashed line indicates the two identical throats for $k(r_m)=0$.
  }
\label{fig:throat3}
\end{figure}
%%%%%%%%%%%%%%%%%%%%%%%%%%%%%%%%%%%%%%%%%%%%%%%%%%%%%%%%%%%%%%%%%%%%%%

When both $z_{\rm throat1,2}$ and $g(r_m)$ are nonzero, two throats are no longer identical  and the geometry  deviates from the warped CY$_3$.
In particular, since two throats contain the different amounts of the NSNS 3-form flux,  $K_0+k(r_m)$ for throat 1 and $K_0-k(r_m)$ for   throat 2, respectively, we expect that two throat lengths would be different (here $K=2K_0+K_t$ as considered in Sec. \ref{Sec:geo-double-throat}).
 We can see this as follows.
  Let $r_m$ be the position of the tip of throat 1 for $k(r_m)=0$.
 As we have seen, this is the distance between the position of the   tip for $\epsilon = 0$ and that for $\epsilon \ne0$ in the deformed throat.
 If $k(r_m)=0$, two throats are identical, so the corresponding distance for throat 2 is also given by $r_m$.
 For a nonzero $k(r_m)$, on the other hand, the different $z$ values    given by \eqref{eq:thraxzs} result in the different $r_m$ values for two throats, 
  \dis{r_{m1}=e^{-\frac{2\pi}{3g_sM}k(r_m)}r_{m},\quad
 r_{m2}=e^{\frac{2\pi}{3g_sM}k(r_m)}r_{m},\label{eq:thrrcoor}}
  as can be inferred from \eqref{eq:r/tau} (see Fig. \ref{fig:throat3}).
Here, for simplicity, we set $m'(r_m)=0$ such that  $z_{\rm throat1}$ and $z_{\rm throat2}$  have the same phase.
Such different $r_m$ values  reflect  the different amounts of the flux accumulated at two tips.
Moreover, $r_{c1}$, the  distance to the UV end of throat 1 from the position of the tip of throat 1 for $\epsilon=0$, and $r_{c2}$, the corresponding distance for throat 2, remain unchanged from the original value $r_c$ to give the correct amounts of the flux contained in two throats.
That is, since $r_{c1}=r_{c2}=r_c$, the NSNS 3-form flux contained in throat 1 is given by
\dis{\frac{3g_sM}{2\pi}\log\Big(\frac{r_{c1}}{r_{m1}}\Big)=\frac{3g_sM}{2\pi}\log\Big(\frac{r_{c}}{e^{-\frac{2\pi}{3g_sM}k(r_m)}r_{m}}\Big)=K_0 + k(r_m),}
whereas the flux contained in throat 2 is,
\dis{\frac{3g_sM}{2\pi}\log\Big(\frac{r_{c1}}{r_{m2}}\Big)=\frac{3g_sM}{2\pi}\log\Big(\frac{r_{c}}{e^{\frac{2\pi}{3g_sM}k(r_m)}r_{m}}\Big)=K_0 - k(r_m),}
as expected.
Then  the ratio between two throat lengths becomes
\dis{\frac{r_{c2}-r_{m2}}{r_{c1}-r_{m1}}=\frac{r_c-e^{\frac{2\pi}{3g_sM}k(r_m)}r_{m}}{r_c-e^{-\frac{2\pi}{3g_sM}k(r_m)}r_{m}},}
which is smaller than one for $k(r_m)>0$.

In this way, in the thraxion scenario, the  different $z$ values at two throat tips are responsible for the different amounts of the NSNS 3-form flux contained in two throats,  making geometries of two throats different.
The similar situation also can be found in the double-throat system with $\overline{\rm D3}$-branes at the tip of  throat 1 only, as extensively discussed in Sec. \ref{sec:B/Fann}.
In both cases, the geometry deviates from the warped CY$_3$ by the different distributions of the NSNS 3-form flux in two throats.
They result from the different local dynamics at two throat   tips  and irrelevant to the dynamics far from the tips.  
  In this regard,  our description of the throat 1 geometry in Sec. \ref{Sec:geo-double-throat} may seem unnatural at first glance as  the additional flux  $p/M$ in throat 1 enhances $r_c$, the position of the UV end, by the factor ${\rm exp}[\frac{2\pi}{3g_sM}\frac{p}{M}]$ as \eqref{eq:ridentity}, whereas $r_m$, the position of the tip, just gets corrected by the backreaction of $\overline{\rm D3}$-branes as \eqref{eq:zratio}.
  This indeed comes from the fact that, $z$ in our discussion depends only on the {\it total} NSNS 3-form flux quantum $K$ following \eqref{eq:K/W}, irrelevant to the distribution of the flux.
  The additional flux $p/M$ in throat 1 is already contained in $K$, hence in $r_m$.
  The factor ${\rm exp}[\frac{2\pi}{3g_sM}\frac{p}{M}]$ in \eqref{eq:ridentity} means that away from the backreaction of $\overline{\rm D3}$-branes giving $r_m/r_{m0}$, the position of the UV end of throat 1 is not changed compared to that before putting $p$  $\overline{\rm D3}$-branes, through the cancellation with the factor ${\rm exp}[-\frac{2\pi}{3g_sM}\frac{p}{M}]$ already contained in $r_{c0}$.
 That is, as can be found in Sec. \ref{Sec:geo-double-throat}, whereas throat 2 contains the smaller amount of the flux $K_2=K_1-(p/M)$ compared to throat 1,    $z_{\rm throat2}$ depends only on the total flux quantum $K$, just like $z_{\rm throat1}$.
 The difference between $z_{\rm throat1}$ and $z_{\rm throat2}$ is a consequence of the backreaction of $\overline{\rm D3}$-branes on the tip of throat 1 only, not of the different amounts of flux contained in two throats.
 Thus $r_{c0}$ is proportional to $z_{\rm throat2}^{1/3}= {\rm exp}[-\frac{2\pi}{3g_sM}K]$, in which $K$ already contains $p/M$.
 When the total NSNS 3-form flux quantum is changed through the brane/flux annihilation as  discussed in Sec. \ref{sec:dthroatb/f}, not only $z_{\rm throat1}$ but also $z_{\rm throat2}$ is  changed  even though any dynamics of branes or fluxes takes place in throat 1.
 This results in the change in the $r$ coordinate in throat 2, not just that in throat 1 through \eqref{eq:r/tau}.
 We can see this explicitly by rescaling the $r$ coordinate in throat 2 to coincide with the $r$ coordinate in the absence of $\overline{\rm D3}$-branes (thus  $p=0=\psi$), such that $r^\psi_{m0}$ and $r^\psi_{c0}$ in \eqref{eq:throat2coor} are transformed to
 \dis{{r'}^{\psi}_{m0}=e^{\frac{2\pi}{3g_sM}\big(\frac{p}{M}-\big\lceil \frac{1}{M}(p-v(\psi)) \big\rceil\big)}r_{m0}^\psi,\quad\quad
 {r'}^{\psi}_{c0}=e^{\frac{2\pi}{3g_sM}\big(\frac{p}{M}-\big\lceil \frac{1}{M}(p-v(\psi)) \big\rceil \big)}r_{c0}^\psi.}
 In terms of them, the positions of the tip and the UV end of   throat 1  given by \eqref{eq:throat1coor} can be rewritten as
 \dis{r_m^\psi=e^{-\frac{2\pi}{3g_sM}\big(\frac{p}{M}-\big\lceil \frac{1}{M}(p-v(\psi)) \big\rceil\big)-\frac13\gamma(v(\psi), 0)}{r'}^{\psi}_{m0},\quad\quad
 r_c^\psi=e^{-\frac13\gamma(v(\psi), 0)}{r'}^{\psi}_{c0}.}
 That is, if the $r$ coordinate in throat 2 were to remain unchanged by the dynamics of $\overline{\rm D3}$-branes at the tip of throat 1, the positions of the tip and the UV end of throat 2 would be given by ${r'}^{\psi}_{m0}$ and  ${r'}^{\psi}_{c0}$, respectively. %, instead of ${r}^{\psi}_{m0}$ and  ${r}^{\psi}_{c0}$.
 Then  the relation between $r^\psi_m$ and ${r'}^{\psi}_{m0}$ is the similar to that between $r_{m1}$ and $r_m$ in the thraxion scenario given by \eqref{eq:thrrcoor}, which is interpreted as the accumulation of the flux at the tip of throat 1.
 However, in our description of the brane/flux annihilation, the $r$ coordinate is changed in throat 2 as well as in throat 1.
 This is because the $r$ coordinate depends on $z$  which is controlled by   the total NSNS 3-form quantum only, without being affected by the   flux distributions in two throats.
 
 We now try to describe $k(r_m)$, the flux accumulated at the tips, appearing in \eqref{eq:thraxzs} in the thraxion scenario using $z_{\rm throat1,2}={\rm exp}[-\frac{2\pi K}{g_sM}]$  in our description of the brane/flux annihilation, which  just depend on the total flux quantum.
 For this purpose,  we recall that the flux distribution is a function of  the $r$ coordinate, which   is proportional to a combination $(z e^y)^{1/3}$ as given by \eqref{eq:r/tau}.
 Then \eqref{eq:thrrcoor} can be rewritten in another way :
 \dis{&r_{m1}=\Big(e^{-\frac{2\pi}{g_sM}(K+k(r_m))}e^y\Big)^{1/3}=\Big(e^{-\frac{2\pi}{g_sM}K}e^{y-\frac{2\pi}{g_sM}k(r_m)}\Big)^{1/3},
 \\
 &r_{m2}=\Big(e^{-\frac{2\pi}{g_sM}(K-k(r_m))}e^y\Big)^{1/3}=\Big(e^{-\frac{2\pi}{g_sM}K}e^{y+\frac{2\pi}{g_sM}k(r_m)}\Big)^{1/3}.}
 Here the last expressions may be interpreted as the shifts in the $y$ coordinate at throat tips.
  This indeed is consistent with the nature of the thraxion, the parametrization of the $S^2$ resolution   at the tip, since  the shift $y \to y+y_0$ modifies the $S^2$ part of the metric in \eqref{eq:tipmetric} from $(y^2/4)(d\tilde{\omega}^2+\sin^2\tilde{\omega}d\tilde{\varphi}^2)$ to $((y+y_0)^2/4)(d\tilde{\omega}^2+\sin^2\tilde{\omega}d\tilde{\varphi}^2)$ such that $S^2$ does not shrink even at the tip, $y=0$. 
 Suppose  $k(r_m)$ is positive.
 Then for throat 2, $y_0=\frac{2\pi}{g_sM}k(r_m)$ is positive hence $y+y_0$, the size of $S^2$,   is always  positive and monotonically increasing as we move away from the tip.
 In contrast, for throat 1, $y_0=-\frac{2\pi}{g_sM}k(r_m)$ is negative hence $y+y_0 <0$ for $0\leq y<|y_0|$.
 Nevertheless, since  $y$ appears in the transformed   metric   in the form of $(y+y_0)^2$, the absolute value of $y+y_0$ is physically meaningful.
 Then  the negative $y_0$ may describe the situation  that even though  the size of $S^2$ is given by nonzero $|y_0|$ at the tip ($y=0$), it collapses at $y=|y_0|$ beyond which  monotonically increases  as $r$ increases.

 \section{Summary and discussion}
\label{sec:conclusion}

 In this article, we investigate the brane/flux annihilation in the double throat system, in which the corresponding cycles of two throats are homologically equivalent.
 Whereas  properties of the CY$_3$ geometry require that two throats have the same geometry, putting $\overline{\rm D3}$-branes at only one of the throat tips  breaks the equivalence, thus the geometry  deviates from the warped CY$_3$.
 The similar geometry also can be found in the recently proposed thraxion scenario.
 As the  brane/flux annihilation goes on, the number of $\overline{\rm D3}$-branes decreases, and the state becomes more supersymmetric.
 Especially, if $p/M$ is an integer, the D3-brane number at the final stage vanishes completely, which indicates that the equivalence of two throats is recovered.

 The dynamics of  the brane/flux annihilation shows that  the stability of the flux distribution  depends on whether the corresponding state is supersymmetric.
 The additional  $\overline{\rm D3}$-branes at the throat tip(s) breaks SUSY.
  At the same time,  the geometry deviates from  CY$_3$, which is reflected in the flux distribution.
 Since the state in this case is unstable, the transition toward more stable (or equivalently, more supersymmetric) state occurs through the brane/flux annihilation.
 Then the geometry becomes close to CY$_3$, which can be described by the change in the flux distribution.
  When two throats are homologically equivalent but $\overline{\rm D3}$-branes are located at only one of throat tips, the corresponding state obviously breaks SUSY as the geometry deviates from CY$_3$.
 Then the flux distributions in two throat regions are different and  the brane/flux annihilation  can be thought of as the process for recovering the identical flux distribution.
 The same feature also can be found in the thraxion scenario.
  The thraxion  is the promotion of the difference between the flux distributions of two throats to a complex scalar.
  It parametrizes the deviation of the geometry from CY$_3$, or equivalently, the instability of the state by   SUSY breaking, as  the different flux distributions do in the brane/flux annihilation.
Thus we also expect that the thraxion takes nonzero value for the SUSY breaking state and would be stabilized in the direction of restoring SUSY, or the identical flux distribution.
  Presumably, we may understand  the features and   constraints of the flux distribution determined by the explicit thraxion potential by comparing with the  brane/flux annihilation and concerning SUSY, which is the subject of the future work.

 In any case, the excursion of thraxion along the potential is expected to be a good candidate for the inflation model, as it shares the number of features with the axion monodromy realization of the brane/flux annihilation.
 Such a resemblance may suggest that even though the brane/flux annihilation is a process local to the throat tip containing the $\overline{\rm D3}$-branes, thus the local observer cannot find the existence of another throat, the inflation model constructed from the brane/flux annihilation has the equivalent description based on thraxion scenario, which assumes the double-throat system.
 % In particular,  since the brane/flux annihilation is a dynamic process which can be applied to the axion monodromy, we expect that various features of  the axion monodromy in this case   also appear in the thraxion scenario. 
%   For instance, the nonzero NSNS 3-form flux at the throat tips in the thraxion scenario can backreact on the geometry, which eventually restricts the viable thraxion field range, through, for instance, changing the value of $|z|$ in the form of \eqref{eq:zmodified}.
 % Presumably, the dynamical stabilization of the thraxion is described in the   similar way to the brane/flux annihilation.
   Meanwhile,    the inflation models based on the dynamics of branes on the throat such as DBI   model \cite{Silverstein:2003hf} (see, e.g., \cite{Kecskemeti:2006cg, McAllister:2007bg, Chen:2008hz, Seo:2018abc, Mizuno:2019pcm} for relevant discussions)  can be extended to the double-throat case, which may reveal nontrivial features of the throat dynamics.

\subsection*{Acknowledgements}

MS is grateful to Kang-Sin Choi and Pablo Soler for discussions and comments    while this work was under progress.
%

%

%\newpage

\appendix

\renewcommand{\theequation}{\Alph{section}.\arabic{equation}}

%\section{Uncertainty for the infrared modes}
%\label{app:IRuncert}
%\setcounter{equation}{0}

\end{document}